\documentclass[aps,prl,twocolumn,floatfix,superscriptaddress]{revtex4}
\pdfoutput=1
\usepackage{amsfonts,amsmath,color,graphicx,subfigure}
\usepackage{url}
\usepackage{subfigure}

\newcommand{\Pe}{{\text{Pe}}}

\begin{document}

\title{Optimization in First-Passage Resetting}
\author{B. De Bruyne}\email{benjamin.debruyne@centraliens.net}
\affiliation{Perimeter Institute, 31 Caroline Street North, Waterloo, ON, N2L
  2Y5, Canada}
\affiliation{CentraleSup\'elec, Universit\'e Paris-Saclay, 3 rue Joliot-Curie, 91190  Gif-sur-Yvette, France}
\author{J Randon-Furling}\email{j.randon-furling@cantab.net}
\affiliation{SAMM, Universit\'e Paris 1 Panth\'eon Sorbonne -- FP2M (FR2036) CNRS, 90 rue de Tolbiac,
75013 Paris, France}
\affiliation{Department of Mathematics, Columbia University, 2990 Broadway, New York, NY 10027, USA}
\author{S. Redner}\email{redner@santafe.edu}
\affiliation{Santa Fe Institute, 1399 Hyde Park Rd., Santa Fe, New Mexico 87501, USA}

\begin{abstract}
  We investigate classic diffusion with the added feature that a diffusing
  particle is reset to its starting point each time the particle reaches a
  specified threshold. In an infinite domain, this process is non-stationary
  and its probability distribution exhibits rich features. In a finite
  domain, we define a non-trivial optimization in which a cost is incurred
  whenever the particle is reset and a reward is obtained while the particle
  stays near the reset point.  We derive the condition to optimize the net
  gain in this system, namely, the reward minus the cost.
  
\end{abstract}

\maketitle 

Diffusion is a fundamental process underlying a wide variety of stochastic
phenomena that has broad applications to physics, chemistry, finance, and
social
sciences~\cite{chandrasekhar1943stochastic,van1992stochastic,chicheportiche2014some,rogers2010diffusion}.
A fruitful recent development is the notion of \emph{resetting}, in which a
diffusing particle is reset to its starting point at a specified
rate~\cite{evans2011diffusion,evans2011diffusionjpa,evans2020stochastic}.
Resetting alters the diffusive motion in fundamental ways and has sparked much
research on its rich consequences (see, e.g.,
\cite{PhysRevLett.112.240601,christou2015diffusion,PhysRevE.92.060101,PhysRevE.92.052126,reuveni2016optimal,pal2017first,belan2018restart,bodrova2019nonrenewal}).
Resetting also has natural applications to search processes, where the search
begins anew if the target is not found within a certain
time~\cite{PhysRevE.92.052127,PhysRevE.92.062115,bhat2016stochastic,eule2016non}.
For such diffusive searches, resetting leads to a dramatic effect: an
infinite search time to find a target becomes finite, with the search time
minimized at a critical reset rate. 

\begin{figure}[ht]
\center{
\includegraphics[width=0.35\textwidth]{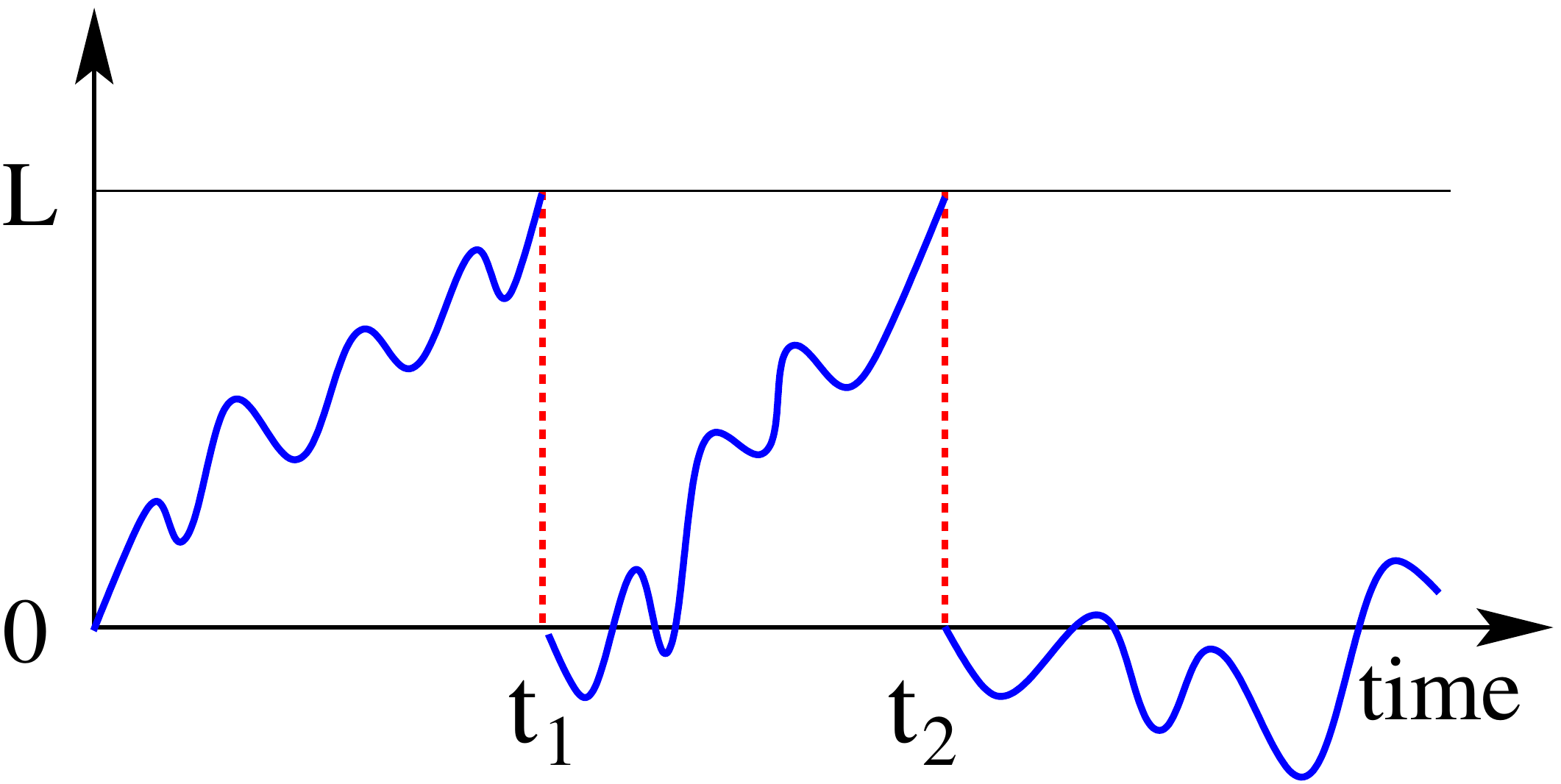}}
\caption{First-passage resetting on the semi-infinite line.  Whenever a
  diffusing particle, which starts at the origin, reaches $L$, it is
  instantaneously reset to the origin.  Successive first-passage times are
  denoted by $t_1, t_2,\ldots$.}
\label{fig:model}
\end{figure}

In this work, we investigate \emph{first-passage resetting} in which
resetting occurs whenever a diffusing particle reaches a threshold location
(Fig.~\ref{fig:model}).  Hence the time of the reset event is determined by
the state of the system itself rather than being imposed
externally~\cite{evans2011diffusion,evans2011diffusionjpa,evans2020stochastic}. First-passage
resetting typifies \emph{regenerative} processes that are reset at
\textit{renewals}~\cite{ross2014introduction}, which has natural applications
to reliability theory~\cite{gnedenko2014mathematical}.  This mechanism was
first envisaged by Feller~\cite{feller1954diffusion} who proved existence and
uniqueness theorems.  Similar ideas were pursued
in~\cite{sherman1958limiting}, and they first appear in the physics
literature in~\cite{falcao2017interacting}. Specifically,
\cite{falcao2017interacting} examines two Brownian particles biased toward
each other that reset to their initial positions upon encounter.  This
corresponds to a drift toward the origin in our semi-infinite geometry of
Fig.~\ref{fig:model}. This negative drift leads to a stationary state, but
the absence of drift leads to a variety of new phenomena, as discussed
below. Moreover, we introduce a path decomposition that provides the spatial
probability distribution in a geometric way.

When the diffusing particle is confined to the finite interval $[0,L]$, we
define an optimization problem in which there is a cost for each resetting
event and an increasing reward as the particle approaches the resetting point
$x=L$. This scenario is inspired by a power-management
problem~\cite{317656,867175}, where the power delivered corresponds to the
coordinate $x$ and a blackout corresponds to resetting. A closely related
optimization arises in finance~\cite{miller1966model}. In both examples, one
seeks to operate almost to full capacity while avoiding saturation: in our
optimization framework, the goal is to maximize the net gain---the difference
between the reward and the cost.

\smallskip

\noindent \emph{Semi-Infinite Geometry:} We first treat diffusion on the
semi-infinite line with first-passage resetting (Fig.~\ref{fig:model}).  The
particle starts at $(x,t)=(0,0)$ and diffuses in the range $x<L$.  When $L$
is reached, the particle is instantaneously reset to the origin.  Define
$F_n(L,t)$ as the probability that the particle resets for the $n^{\text th}$
time at time $t$.  For $n=1$, this quantity is the first-passage probability
for a diffusing particle to reach $L$~\cite{redner2001guide}
\begin{align*}
F_1(L,t) = \frac{L}{\sqrt{4\pi D t^3}}\, e^{-L^2/4Dt}\,.
\end{align*}
For the particle to reset for the $n^{\text{th}}$ time at time $t$, it must
reset for the $(n-1)^{\text{th}}$ time at time $t'<t$, and reset one more
time at time $t$.  Because the process is renewed at each reset, $F_n(L,t)$
is given by the renewal equation
\begin{align}
\label{eq:Fnsi}
  F_n(L,t) = \int_0^t dt'\, F_{n-1}(L,t')\, F_1(L,t-t'),\quad n>1\,.
\end{align}
The convolution structure of Eq.~\eqref{eq:Fnsi} lends itself to a Laplace
transform analysis because the corresponding equation in the Laplace domain
is simply $\widetilde F_n(L,s)=\widetilde F_{n-1}(L,s) \widetilde F_1(L,s)$
from which $\widetilde F_n(L,s)= \widetilde F_1(L,s)^n$.

Using the Laplace transform of the first-passage probability
$\widetilde F_1(L,s)= e^{-y_{\scriptscriptstyle L}}$, we thus obtain
$\widetilde F_n(L,s)= e^{-ny_{\scriptscriptstyle L}}$, where we introduce
$y=x\sqrt{s/D}$ and $y_{\scriptscriptstyle L}=L\sqrt{s/D}$ for notational
simplicity.  Notice that $\widetilde F_n(L,s)$ has the same form as
$\widetilde F_1(L,s)$ with $L\to nL$.  That is, the time for a diffusing
particle to reset $n$ times is the same as the time for a freely diffusing
particle to first reach $nL$.

In contrast to fixed-rate resetting, the spatial probability distribution in
the semi-infinite geometry is \emph{non-stationary}.  This distribution is
formally determined by
\begin{subequations}
  \label{eq:renEq}
\begin{align}
  \label{Pxt}
    P(x,t) = G(x,L,t)+
     \sum_{n\geq 1} \int_0^{t}\! dt '\, F_n(L,t')\,
  G(x,L,t\!-\!t'),
\end{align}
with $G(x,L,t) =\big[e^{-x^2/4Dt} - e^{-(x-2L)^2/4Dt}\big]/\sqrt{4\pi Dt}$,
the probability for a particle to be at $(x,t)$ when it starts at the origin
in the presence of an absorbing boundary at
$x=L$~\cite{feller2008introduction,redner2001guide,Bray13}.  Equation~\eqref{Pxt}
states that for the particle to be at $(x,t)$, it either (i) must never
hit~$L$, in which case its probability distribution is just $G(x,L,t)$, or
(ii), the particle first hits~$L$ for the $n^{\rm th}$ time at $t'<t$, after
which the particle restarts at the origin and then propagates to $x$ in the
remaining time $t-t'$ without hitting~$L$ again. The latter set of
trajectories must be summed over all~$n$.  An equivalent way of writing
Eq.~\eqref{Pxt} is
\begin{align}
\label{eq:Pxtb}
P(x,t) =  G(x,L,t) + \int_0^t dt' \,F_{1}(L,t')P(x,t-t')\,.
\end{align}
\end{subequations}
The first term accounts for the particle never reaching $x=L$, while the
second term accounts for the particle reaching $x=L$ at time $t'$, after
which the process starts anew from $(x,t)=(0,t')$ for the remaining time
$t-t'$.  Analogously to the Fokker-Planck equations, we refer to
Eqs.~\eqref{Pxt} and \eqref{eq:Pxtb} as the \emph{forward} and
\emph{backward} renewal equations, respectively.

To solve for $P(x,t)$ we again treat the problem in the Laplace domain.
While we can find the solution from the Laplace transform of Eq.~\eqref{Pxt},
the solution is simpler and more direct from the Laplace transform
of~\eqref{eq:Pxtb}:
\begin{align}
 \label{Pxs-sol}
  \widetilde{P}(y,s) &=   \frac{\widetilde{G}(y,y_{\scriptscriptstyle L},s)}
        {1-\widetilde{F}_1(y_{\scriptscriptstyle L},s)}
        =  \frac{1}{\sqrt{4Ds}}\,\frac{\left[
  e^{-|y|} -  e^{-|y-2y_{\scriptscriptstyle L}|}\right]}
 {1-e^{-y_{\scriptscriptstyle L}}}\,.
\end{align}
We now need to separately consider the cases
$0\leq y\leq y_{\scriptscriptstyle L}$ and $y<0$.  In the former case, we
expand the denominator in a Taylor series to give
\begin{subequations}
\label{P<}
  \begin{align}
 \label{Pxs-sol<}
  \widetilde{P}(y,s)
    & = \frac{1}{\sqrt{4Ds}}\,
  \left[ e^{-y}-e^{-(2y_{\scriptscriptstyle L}-y)}\right]\,
      \sum_{n\geq 0} e^{-ny_{\scriptscriptstyle L}}\nonumber\\
    &=  \frac{1}{\sqrt{4Ds}}\,\sum_{n\geq 0}
      \left[ e^{-(y+ny_{\scriptscriptstyle L})}
      -e^{-[(n+2)y_{\scriptscriptstyle L}-y]}\right]\,,
\end{align}
from which 
\begin{align}
 \label{Pxt-sol<}
  P(x,t)
  &\!=\!  \frac{1}{\sqrt{4\pi Dt}}\,\sum_{n\geq 0}\left[ e^{-(x\!+\!nL)^2/4Dt}
    \!-\!e^{-[x-(n\!+\!2)L]^2/4Dt}\right].
\end{align}
\end{subequations}
The long-time limit of $P(x,t)$ is particularly simple.  By expanding the
Laplace transform for small $s$ and then inverting this transform, we find
\begin{align*}
  P(x,t)\simeq \frac{1}{\sqrt{\pi Dt}}\, \frac{L-x}{L}\qquad 0\leq
  x\leq L\,.
\end{align*}
This linear form arises from the balance of the diffusive flux exiting at
$x=L$ that is re-injected at $x=0$.

\begin{figure}[ht]
\subfigure[]{\includegraphics[width=0.23\textwidth]{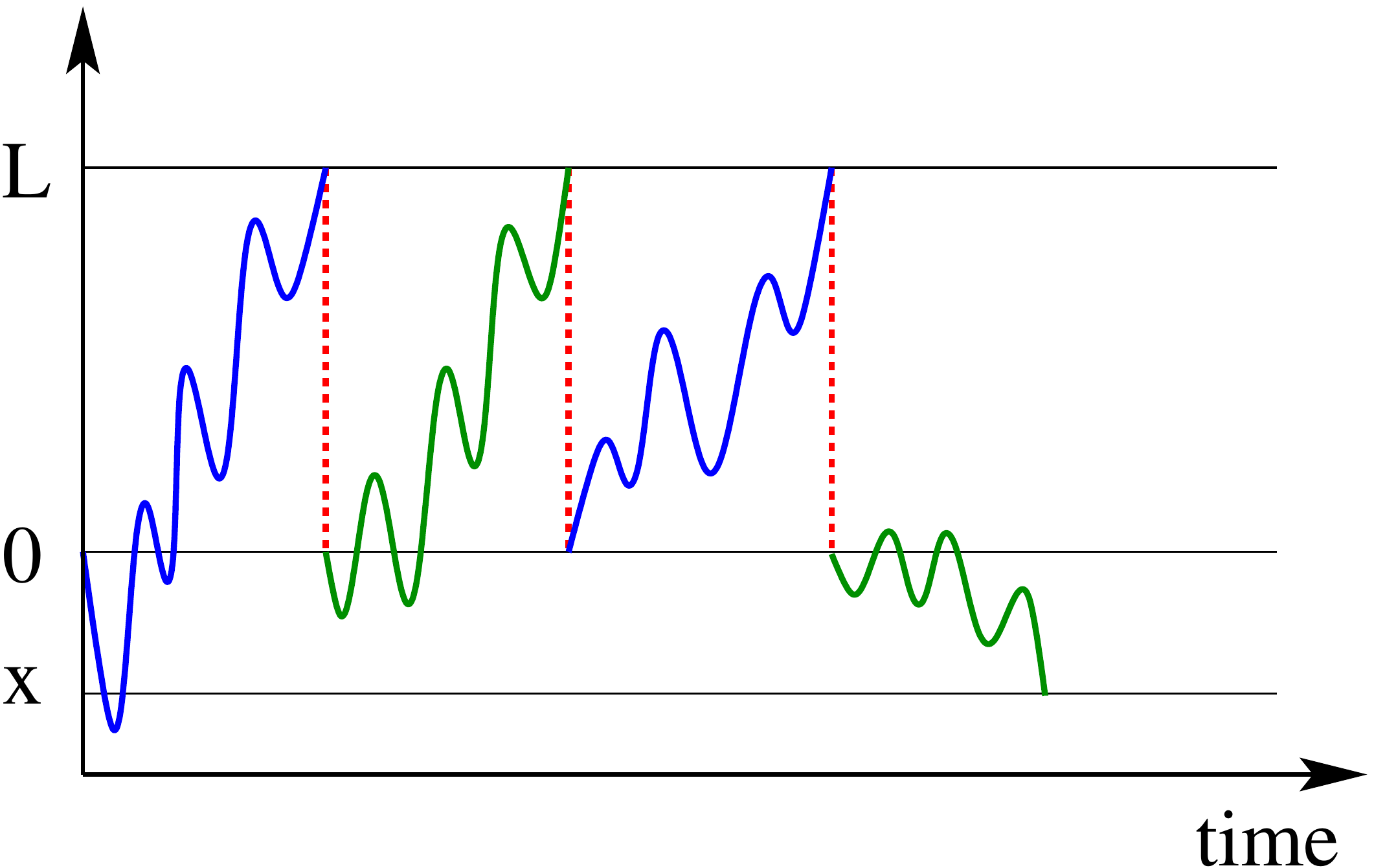}}\quad
\subfigure[]{\includegraphics[width=0.23\textwidth]{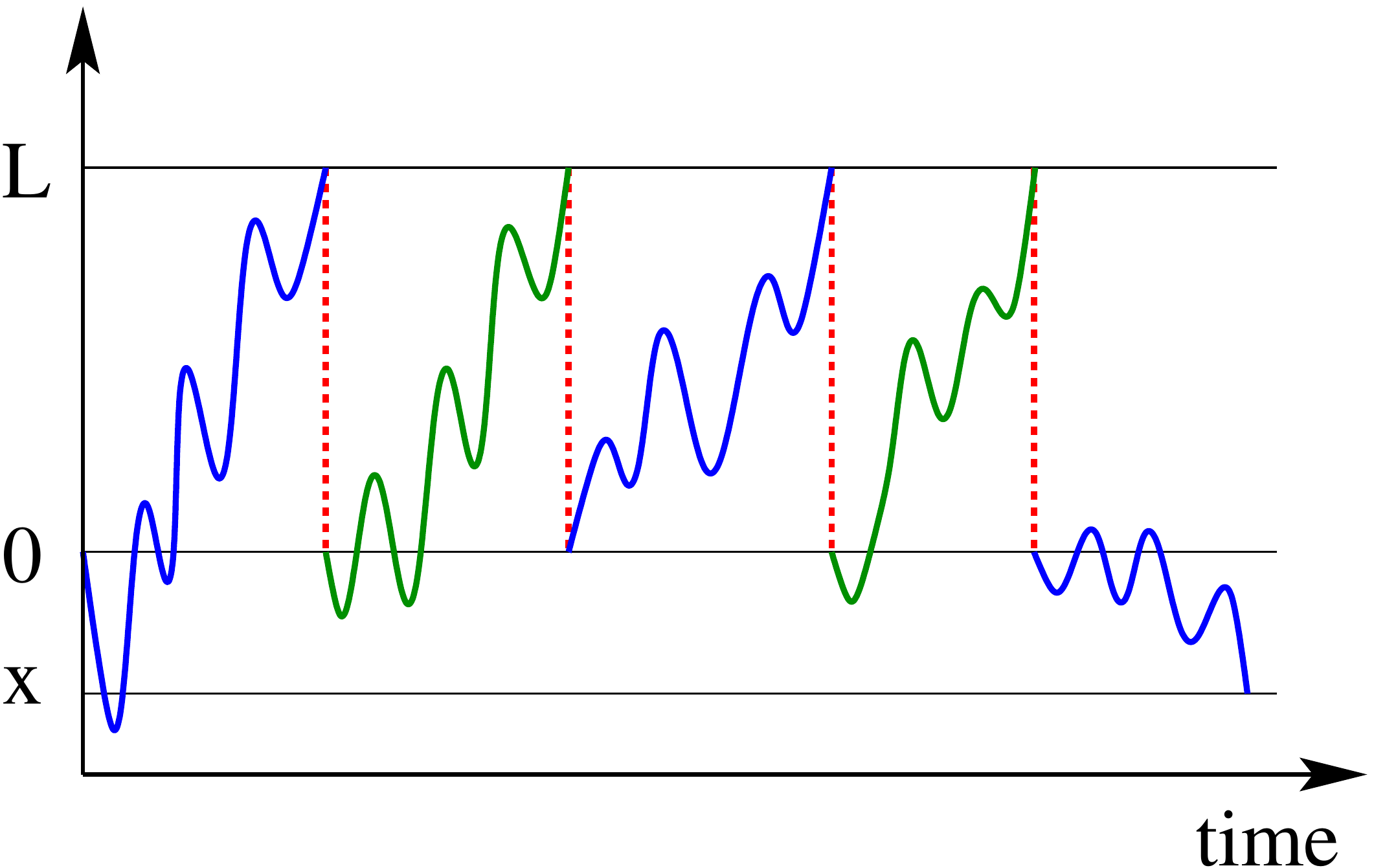}}\\
\subfigure[]{\includegraphics[width=0.23\textwidth]{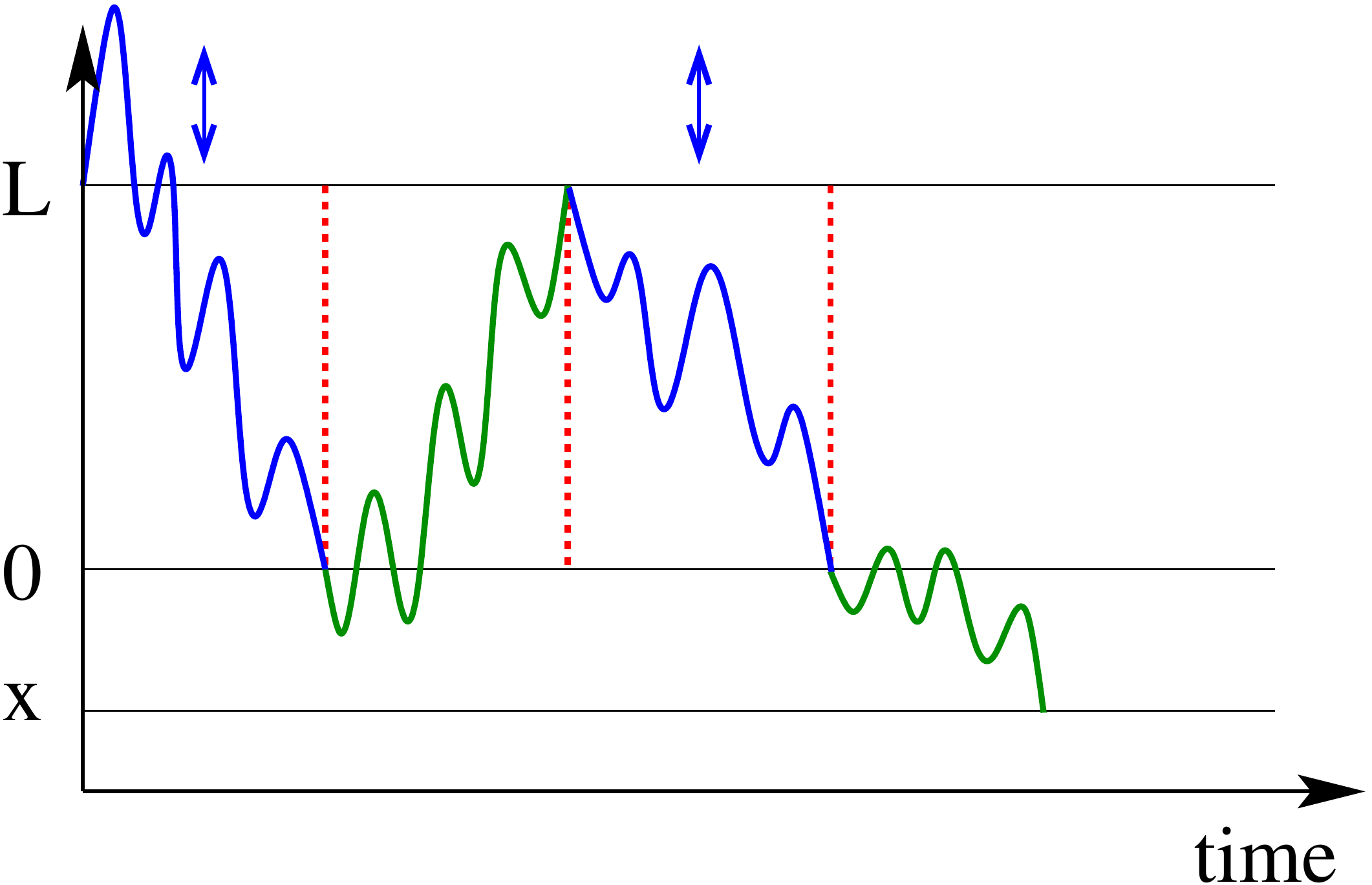}}\quad
\subfigure[]{\includegraphics[width=0.23\textwidth]{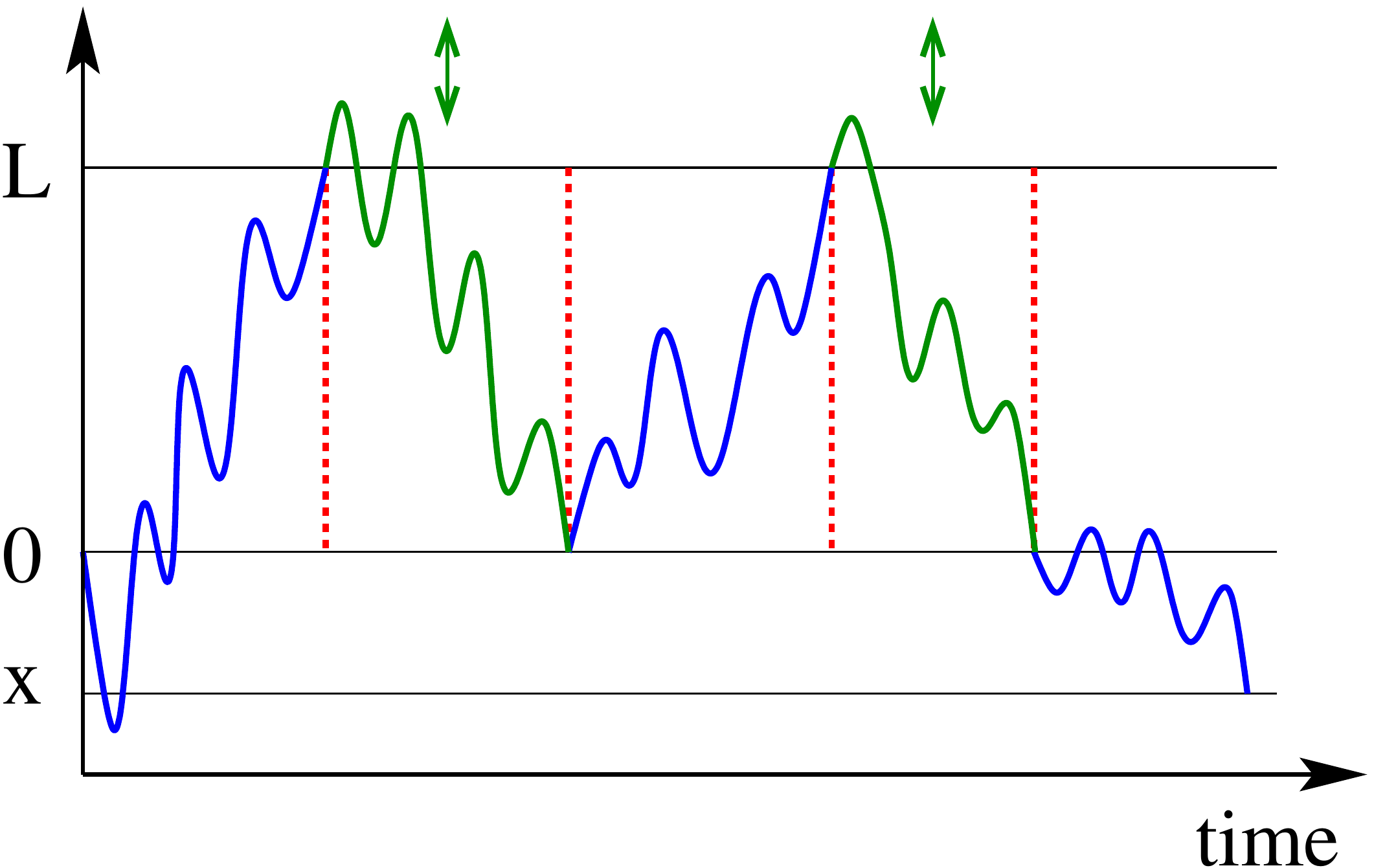}}
\caption{Schematic of diffusion with first-passage resetting.  Shown are
  paths from $(0,0)$ to $(x,t)$ with: (a) an odd number or (b) an even number
  of reset events.  We transform either the odd-numbered or the even-numbered
  pieces of the original path through the reflection $x(t)\rightarrow L-x(t)$
  to yield free diffusion from either: (c) $(L,0)$ to $(x,t)$, or (d) $(0,0)$
  to $(x,t)$. Summing over all numbers of reset events gives
  Eq.~\eqref{Pxt<}.}
    \label{fig:pathtsfm}
\end{figure}

For $y<0$, $\widetilde{P}(y,s)$ in Eq.~\eqref{Pxs-sol} is factorizable:
\begin{subequations}
\begin{align}
  \widetilde{P}(y,s) &= \frac{1}{\sqrt{4Ds}}\!\left[\frac{e^{y}\!-\!
  e^{y-2y_{\scriptscriptstyle L}}}{1-e^{-y_{\scriptscriptstyle L}}}\right]
  =  \frac{1}{\sqrt{4Ds}}\!\left[e^{y}\!+\! e^{(y-y_{\scriptscriptstyle L})}\right],
\end{align}
and this latter form can be readily inverted to give
\begin{align}
  \label{Pxt<}
  P(x,t) = \frac{1}{\sqrt{4\pi Dt}}\,
  \left[e^{-x^2/4Dt}+ e^{-(x-L)^2/4Dt}\right]\quad x<0\,.
\end{align}
\end{subequations}
Strikingly, this closed form represents the superposition of free diffusion
paths to $(x,t)$ starting from $(0,0)$ and from $(L,0)$.  This property may
be derived by decomposing trajectories with resetting into a series of
first-passage segments between each reset and then reflecting and translating
them to obtain either a free diffusion path that starts at $(L,0)$ or at
$(0,0)$ and propagates to $(x,t)$ as indicated in Fig.~\ref{fig:pathtsfm}.
We emphasize that this decomposition applies for any symmetric stochastic
process.

A basic characteristic of regenerative processes is the number of reset
events up to time $t$.  The probability $Q_n$ for $n$ reset events is given
by
\begin{align*}
Q_n &= \text{Prob}(\geq n\ \text{resets})- \text{Prob}(\geq n\!+\!1\ \text{resets})\\
&= \int_0^{t}\! dt '\, \big[F_n(L,t')- F_{n+1}(L,t')\big]\,.
\end{align*}
From our previous result that $F_n(L,t)=F_1(nL,t)$, we immediately have
\begin{equation}
\label{eq:ndist}
Q_n=\mathrm{erf}\left(\frac{(n+1)L}{\sqrt{4Dt}}\right)-\mathrm{erf}\left(\frac{nL}{\sqrt{4Dt}}\right),
\end{equation}
where $\mathrm{erf}$ is the Gauss error function.

We can compute $\mathcal{N}(t)\equiv \langle n \rangle$, the average number
of reset events from~\eqref{eq:ndist}, but it is quicker to express
$\mathcal{N}(t)$ in terms of a backward renewal equation:
\begin{align}
  \label{eq:navgb}
  \mathcal{N}(t) = \int_0^t dt' \,F_{1}(L,t')\big(1+\mathcal{N}(t-t')\big)\,.
\end{align}
Equation~\eqref{eq:navgb} accounts for the particle first hitting $L$ at any
time $t'<t$, then the process is renewed over the time range $t-t'$, so there
will be on average $1+\mathcal{N}(t-t')$ resets.  Taking the Laplace
transform of~\eqref{eq:navgb} then leads to
\begin{align}
\label{eq:Nsi}
\widetilde{\mathcal{N}}(s) &= \frac{\widetilde{F}_{1}(L,s)}{s\,\big[1-\widetilde{F}_{1}(L,s)\big]} 
   = \frac{e^{-y_{\scriptscriptstyle L}}}{s(1-e^{-y_{\scriptscriptstyle L}})},
\end{align}
from which we extract the long-time behavior of the average number of reset events by
taking the $s\rightarrow 0$ limit.  We thus find
$\mathcal{N}(t) \simeq \sqrt{4Dt/\pi L^2}$.

\smallskip

\noindent \emph{Optimization in the Finite Interval:} Let us now view the
coordinate $x$ as the operating point of a mechanical system or a power grid
with $x\in[0,L]$, and a control mechanism that acts upon $x$ in the form of a drift. It is desirable that the system operates close to the
maximum operating point; that is, $x(t)$ near $L$.  We therefore assign a
reward that is proportional to $x(t)/L$.  On the other hand, when $x$ reaches
$L$ the system breaks down.  Each breakdown incurs a cost $C$, after which
the system is reset to the origin.  The goal is to determine the optimal
operation of this system, for which the objective function
\begin{align}
\label{F}
\mathcal{F} =\,\lim_{T\rightarrow \infty}\,\frac{1}{T}\left[ \frac{1}{L} \int_0^T x(t)\,dt - C \mathcal{N}(T)\right]
\end{align}
is maximized, with $\mathcal{N}(T)$ the number of breakdowns within a time
$T$.

As a simple model, we posit that the coordinate $x$ changes in time according
to diffusion, due to demand fluctuations, with superimposed drift.  From a
practical viewpoint, the drift should drive the system away from the
breakdown point; that is, the control mechanism forestalls breakdowns.
However, optimization arises for either sense of the drift.  Mathematically,
we need to solve the probability distribution of the particle which obeys the
convection diffusion equation
\begin{align}
\label{eq:difft}
  \partial_t c  +v\partial_x c
  =D \partial_{xx}c + \delta(x)(-D\partial_x c+vc)\rvert_{x=L} \,,
 \end{align}
subject to the initial and boundary conditions
\begin{align*}
\begin{cases}{}
    \left(D\partial_x c-vc\right)\rvert_{x=0} = \delta(t) \,,\\
    c(L,t)\, = \,c(x,0) = 0\,.
    \end{cases}
\end{align*}
Here, $c\equiv c(x,t)$ is the probability density, the subscripts denote
partial differentiation, $D$ is the diffusion coefficient, and $v$ is the
drift velocity.  The delta-function term in \eqref{eq:difft} corresponds to
the reinjection of the outgoing flux at $x=L$ to $x=0$, and the initial
condition corresponds to starting the system at $(x,t)=(0,0)$.

This problem can be readily solved in the Laplace domain.  As a preliminary,
we first need to solve the simpler subproblem with no flux reinjection so
that the delta-function term in \eqref{eq:difft} is absent.  In this case,
the concentration is~\cite{BRR21}
\begin{subequations}
  \label{eq:c0F1}
\begin{align}
\label{eq:c0}
  \widetilde{c}_0(x,s) &= \frac{2\, e^{vx/2D} \sinh \left[w (L-x)\right]}{\mathcal{M}}\,,
\end{align}
where the subscript 0 denotes the concentration without flux reinjection,
$\Pe\equiv vL/2D$ is the P\'eclet number, $w = \sqrt{v^2+4Ds}/2D$, and
$\mathcal{M}= 2 D w \cosh (L w)+v \sinh (L w)$.  From $\widetilde{c}_0$,
the Laplace transform of the first-passage probability to $x=L$ is
\begin{align}
\label{eq:F1}
\widetilde{F}_1(L,s) &= \left( -D\partial_x \widetilde{c}_0 + v \widetilde{c}_0 \right)\rvert_{x=L}
 = \frac{2 \,D\, w\, e^{\Pe}}{\mathcal{M}}\,.
\end{align}
\end{subequations}

With reinjection of the outgoing flux, the concentration obeys the
renewal equations \eqref{eq:renEq}. In the Laplace domain and using
$\widetilde{c}_0$ above, we find
\begin{align}
\label{eq:part}
  \widetilde{c}(x,s)= \frac{\widetilde{c}_0(x,s)}{1-\widetilde{F}_1(L,s)}
  = \frac{2\, e^{vx/2D} \sinh [w (L-x)]}{\mathcal{M}-2 D\, w \,e^{\Pe}}\,.
\end{align}
where we substitute in Eqs.~\eqref{eq:c0F1} to obtain the final result.
\smallskip

On a finite interval, diffusion with first-passage resetting is ergodic and
admits a steady
state~\cite{grigorescu2002brownian,leung2008spectral,ben2009ergodic}. In
the $s\to 0$ limit, the coefficient of the term proportional to $\frac{1}{s}$
in $\widetilde{c}(x,s)$ gives the steady-state concentration in the time
domain, which is
\begin{align}
  \label{eq:solt}
 c(x) &\simeq
 \frac{1}{L}\times\frac{1\,-\,e^{-2\Pe(L-x)/L}}{1\,-\,\Pe^{-1}\,e^{-\Pe}\,\sinh\left(\Pe\right)}\,,
\end{align}
from which the normalized first moment is
\begin{align}
\label{eq:xavg}
\frac{\langle x\rangle}{L} =\frac{1}{L}\int_0^L \!\!x\,c(x)\, dx=
 \frac{\left(2\Pe^2-2\Pe+1\right)\,e^{2\Pe}-1}{ 2\Pe\left[(2\Pe-1)\,e^{2\Pe}+1\right]} \,.
\end{align}
 
The average number of reset events $\mathcal{N}$ satisfies the backward renewal
equation \eqref{eq:navgb} and using $\widetilde{F}_1$ from \eqref{eq:F1} we
find
\begin{subequations}
\begin{align}
\label{eq:Nssp}
\widetilde{\mathcal{N}}(s) &=  \frac{2 D\, w\, e^{\Pe}}  {s\,
 \left[\mathcal{M}-2 D w\, e^{\Pe}\right]}\,.
\end{align}
We now extract the long-time behavior for the average number of times that
$x=L$ is reached by taking the limit $s\rightarrow 0$ of
$\widetilde{\mathcal{N}}(s)$ to give
\begin{align}
\label{eq:N}
\mathcal{N}(T) &\simeq \frac{4\Pe^2}{2\Pe-1+ e^{-2\Pe}}\,\,\frac{T}{L^2/D}\,.
\end{align}
\end{subequations}

\begin{figure}
     \centering
     \includegraphics[width=0.35\textwidth]{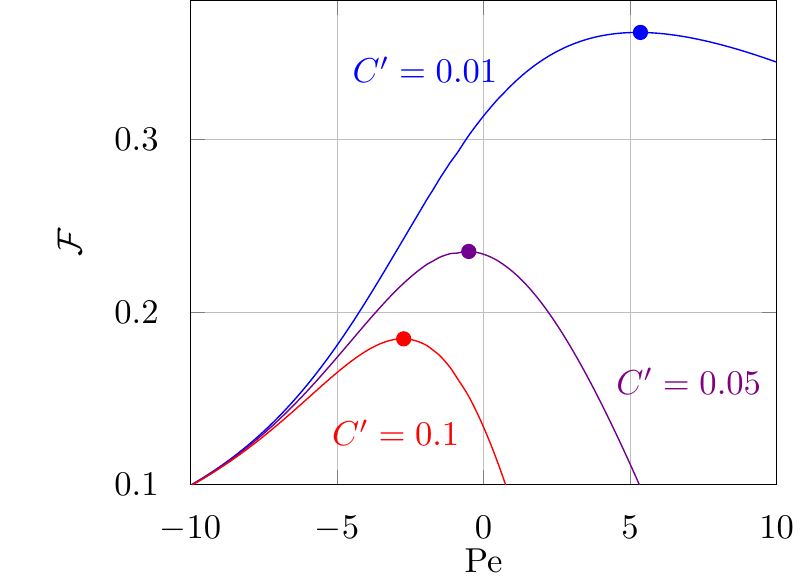}
     \caption{The objective function of Eq.~\eqref{F} versus P\'eclet number $\Pe$ for
       different normalized cost values $C'\equiv C/(L^2/D)$.  Indicated on
       each curve is the optimal operating point.}
     \label{fig:F}
\end{figure}{}

Substituting these expressions for $\langle x\rangle/L$ and $\mathcal{N}$
into \eqref{F} immediately gives the objective function; representative plots
are shown in Fig.~\ref{fig:F}.  The salient feature is that there is an
optimal operating P\'eclet number for each cost value.  When the cost per
breakdown is small, it is advantageous to run the system at positive P\'eclet
number.  Although there are many breakdowns, they are cheap, and there is a
greater reward in pushing the system to the limit.  Conversely, when the
breakdown cost is high, the optimal operating point is at a negative P\'eclet
number.  Although there is little gain in operating the system at such a low
level, the breakdown cost is so high that low-level operation is optimal.

When a mechanical system breaks down, there is downtime when repairs are
effected before the system can be restarted.  Such a delay, akin to the refractory period considered in~\cite{PhysRevE.92.060101,RUR14,reuveni2016optimal,EMref18}, can be
incorporated into our modeling by including a random delay after each
resetting event.  Thus when the particle reaches $x=L$ and is returned to
$x =0$, it waits there a random time $\tau$ that is drawn from the
exponential distribution $\sigma^{-1}e^{-\tau/\sigma}$ before moving again.
Our formalism developed for instantaneous resetting can be naturally extended
to resetting with delay---which might also be viewed as a so-called
``sticky'' Brownian
motion~\cite{gallavotti1972boundary,harrison1981sticky,bou2020sticky}
combined with resetting.  The details are cumbersome, however, and we merely
quote the main results. Upon including delay, the calculational steps that
led to \eqref{eq:xavg} now give~\cite{BRR21}
\begin{align}
\label{eq:xav-d}
\frac{\langle x\rangle}{L}=  \frac{\left[(2\Pe-2)\Pe+1\right]e^{2\Pe}-1}
  {2 \left[\Pe(4\overline{\tau}\,\Pe^2+2\Pe-1)e^{2\Pe}+\Pe\right]} \,,
\end{align}
where $\overline{\tau}=D\sigma /L^2$ is a dimensionless measure of the delay
time. Similarly, following the calculation that led to~\eqref{eq:N}, the
average number of breakdowns is
\begin{align}
  \mathcal{N} =\frac{4\Pe^2}{2\Pe-1+4\overline{\tau}\,\Pe^2+e^{-2\Pe}}\,\frac{T}{L^2/D}\, .
\end{align}
This leads to an objective function whose qualitative features are similar to
the no-delay case.  The primary difference is that the optimal P\'eclet
number and the corresponding optimal objective function $\mathcal{F}$ both
decrease as the delay time is increased.  Indeed, delay reduces the number of
breakdowns but also induces the coordinate to remain closer to the origin.
In the limit where the delay is extremely long, the optimal P\'eclet number
will be small and will almost not depend on the cost per breakdown, as the
particle  almost never hits the resetting boundary. 

In summary, triggering the reset of a diffusing particle by a first-passage
event leads to rich features.  On the infinite line, the probability
distribution is exactly calculable and can be understood in terms of a subtle
path decomposition.  In a finite domain $[0,L]$, there exists an optimal bias
velocity that maintains the system at maximum performance---close to the peak
operating point of $x=L$ while minimizing the number of breakdowns.

The formalism developed here can be extended to systems with multiple degrees
of freedom, such as a power grid, where the breakdown in one coordinate
induces a breakdown in another coordinate.  Another promising direction is to
incorporate the possibility of partial versus complete
repair~\cite{rausand2003system}.  After partial repair, the operating range
of the system is reduced, so that the next breakdown is more likely to occur
sooner.  On the other hand, there will be a smaller penalty associated with
partial repair.  This perspective may allow one to optimize both the
frequency and magnitude of repair costs.

More generally, first-passage resetting may lead to intriguing statistical
features in problems in control theory and management science (where
fluctuations of inventory or cash fund levels are typically modeled by random
walks or Brownian motion, and there generally exists a maximal capacity that
one seeks to use optimally~\cite{harrison1983impulse,buckholtz1983analysis})
or in biology (where allele frequencies in population genetics models evolve
according to diffusion, with ``killing'' when a frequency reaches a limit
level~\cite{tavare1979dual,goel2016stochastic}).
 
\acknowledgments{BBs research at the Perimeter Institute is supported in part
  by the Government of Canada through the Department of Innovation, Science
  and Economic Development Canada and by the Province of Ontario through the
  Ministry of Colleges and Universities. JRFs research at Columbia University
  is supported by the Alliance Program. SR thanks Paul Hines for helpful
  conversations and financial support from NSF grant DMR-1910736.}


\begin{thebibliography}{41}%
\makeatletter
\providecommand \@ifxundefined [1]{%
 \@ifx{#1\undefined}
}%
\providecommand \@ifnum [1]{%
 \ifnum #1\expandafter \@firstoftwo
 \else \expandafter \@secondoftwo
 \fi
}%
\providecommand \@ifx [1]{%
 \ifx #1\expandafter \@firstoftwo
 \else \expandafter \@secondoftwo
 \fi
}%
\providecommand \natexlab [1]{#1}%
\providecommand \enquote  [1]{``#1''}%
\providecommand \bibnamefont  [1]{#1}%
\providecommand \bibfnamefont [1]{#1}%
\providecommand \citenamefont [1]{#1}%
\providecommand \href@noop [0]{\@secondoftwo}%
\providecommand \href [0]{\begingroup \@sanitize@url \@href}%
\providecommand \@href[1]{\@@startlink{#1}\@@href}%
\providecommand \@@href[1]{\endgroup#1\@@endlink}%
\providecommand \@sanitize@url [0]{\catcode `\\12\catcode `\$12\catcode
  `\&12\catcode `\#12\catcode `\^12\catcode `\_12\catcode `\%12\relax}%
\providecommand \@@startlink[1]{}%
\providecommand \@@endlink[0]{}%
\providecommand \url  [0]{\begingroup\@sanitize@url \@url }%
\providecommand \@url [1]{\endgroup\@href {#1}{\urlprefix }}%
\providecommand \urlprefix  [0]{URL }%
\providecommand \Eprint [0]{\href }%
\providecommand \doibase [0]{http://dx.doi.org/}%
\providecommand \selectlanguage [0]{\@gobble}%
\providecommand \bibinfo  [0]{\@secondoftwo}%
\providecommand \bibfield  [0]{\@secondoftwo}%
\providecommand \translation [1]{[#1]}%
\providecommand \BibitemOpen [0]{}%
\providecommand \bibitemStop [0]{}%
\providecommand \bibitemNoStop [0]{.\EOS\space}%
\providecommand \EOS [0]{\spacefactor3000\relax}%
\providecommand \BibitemShut  [1]{\csname bibitem#1\endcsname}%
\let\auto@bib@innerbib\@empty
\bibitem [{\citenamefont {Chandrasekhar}(1943)}]{chandrasekhar1943stochastic}%
  \BibitemOpen
  \bibfield  {author} {\bibinfo {author} {\bibfnamefont {S.}~\bibnamefont
  {Chandrasekhar}},\ }\href@noop {} {\bibfield  {journal} {\bibinfo  {journal}
  {Rev. Mod. Phys.}\ }\textbf {\bibinfo {volume} {15}},\ \bibinfo {pages} {1}
  (\bibinfo {year} {1943})}\BibitemShut {NoStop}%
\bibitem [{\citenamefont {Van~Kampen}(1992)}]{van1992stochastic}%
  \BibitemOpen
  \bibfield  {author} {\bibinfo {author} {\bibfnamefont {N.~G.}\ \bibnamefont
  {Van~Kampen}},\ }\href@noop {} {\emph {\bibinfo {title} {Stochastic processes
  in physics and chemistry}}},\ Vol.~\bibinfo {volume} {1}\ (\bibinfo
  {publisher} {Elsevier},\ \bibinfo {year} {1992})\BibitemShut {NoStop}%
\bibitem [{\citenamefont {Chicheportiche}\ and\ \citenamefont
  {Bouchaud}(2014)}]{chicheportiche2014some}%
  \BibitemOpen
  \bibfield  {author} {\bibinfo {author} {\bibfnamefont {R.}~\bibnamefont
  {Chicheportiche}}\ and\ \bibinfo {author} {\bibfnamefont {J.-P.}\
  \bibnamefont {Bouchaud}},\ }in\ \href@noop {} {\emph {\bibinfo {booktitle}
  {First-passage Phenomena And Their Applications}}}\ (\bibinfo  {publisher}
  {World Scientific},\ \bibinfo {year} {2014})\ pp.\ \bibinfo {pages}
  {447--476}\BibitemShut {NoStop}%
\bibitem [{\citenamefont {Rogers}(2010)}]{rogers2010diffusion}%
  \BibitemOpen
  \bibfield  {author} {\bibinfo {author} {\bibfnamefont {E.~M.}\ \bibnamefont
  {Rogers}},\ }\href@noop {} {\emph {\bibinfo {title} {Diffusion of
  innovations}}}\ (\bibinfo  {publisher} {Simon and Schuster},\ \bibinfo {year}
  {2010})\BibitemShut {NoStop}%
\bibitem [{\citenamefont {Evans}\ and\ \citenamefont
  {Majumdar}(2011{\natexlab{a}})}]{evans2011diffusion}%
  \BibitemOpen
  \bibfield  {author} {\bibinfo {author} {\bibfnamefont {M.~R.}\ \bibnamefont
  {Evans}}\ and\ \bibinfo {author} {\bibfnamefont {S.~N.}\ \bibnamefont
  {Majumdar}},\ }\href@noop {} {\bibfield  {journal} {\bibinfo  {journal}
  {Phys. Rev. Lett.}\ }\textbf {\bibinfo {volume} {106}},\ \bibinfo {pages}
  {160601} (\bibinfo {year} {2011}{\natexlab{a}})}\BibitemShut {NoStop}%
\bibitem [{\citenamefont {Evans}\ and\ \citenamefont
  {Majumdar}(2011{\natexlab{b}})}]{evans2011diffusionjpa}%
  \BibitemOpen
  \bibfield  {author} {\bibinfo {author} {\bibfnamefont {M.~R.}\ \bibnamefont
  {Evans}}\ and\ \bibinfo {author} {\bibfnamefont {S.~N.}\ \bibnamefont
  {Majumdar}},\ }\href@noop {} {\bibfield  {journal} {\bibinfo  {journal} {J.
  Phys. A: Mathematical and Theoretical}\ }\textbf {\bibinfo {volume} {44}},\
  \bibinfo {pages} {435001} (\bibinfo {year} {2011}{\natexlab{b}})}\BibitemShut
  {NoStop}%
\bibitem [{\citenamefont {Evans}\ \emph {et~al.}(2020)\citenamefont {Evans},
  \citenamefont {Majumdar},\ and\ \citenamefont
  {Schehr}}]{evans2020stochastic}%
  \BibitemOpen
  \bibfield  {author} {\bibinfo {author} {\bibfnamefont {M.~R.}\ \bibnamefont
  {Evans}}, \bibinfo {author} {\bibfnamefont {S.~N.}\ \bibnamefont {Majumdar}},
  \ and\ \bibinfo {author} {\bibfnamefont {G.}~\bibnamefont {Schehr}},\
  }\href@noop {} {\bibfield  {journal} {\bibinfo  {journal} {J. Phys. A:
  Mathematical and Theoretical}\ } (\bibinfo {year} {2020})}\BibitemShut
  {NoStop}%
\bibitem [{\citenamefont {Boyer}\ and\ \citenamefont
  {Solis-Salas}(2014)}]{PhysRevLett.112.240601}%
  \BibitemOpen
  \bibfield  {author} {\bibinfo {author} {\bibfnamefont {D.}~\bibnamefont
  {Boyer}}\ and\ \bibinfo {author} {\bibfnamefont {C.}~\bibnamefont
  {Solis-Salas}},\ }\href@noop {} {\bibfield  {journal} {\bibinfo  {journal}
  {Phys. Rev. Lett.}\ }\textbf {\bibinfo {volume} {112}},\ \bibinfo {pages}
  {240601} (\bibinfo {year} {2014})}\BibitemShut {NoStop}%
\bibitem [{\citenamefont {Christou}\ and\ \citenamefont
  {Schadschneider}(2015)}]{christou2015diffusion}%
  \BibitemOpen
  \bibfield  {author} {\bibinfo {author} {\bibfnamefont {C.}~\bibnamefont
  {Christou}}\ and\ \bibinfo {author} {\bibfnamefont {A.}~\bibnamefont
  {Schadschneider}},\ }\href@noop {} {\bibfield  {journal} {\bibinfo  {journal}
  {J. Phys. A: Mathematical and Theoretical}\ }\textbf {\bibinfo {volume}
  {48}},\ \bibinfo {pages} {285003} (\bibinfo {year} {2015})}\BibitemShut
  {NoStop}%
\bibitem [{\citenamefont {Rotbart}\ \emph {et~al.}(2015)\citenamefont
  {Rotbart}, \citenamefont {Reuveni},\ and\ \citenamefont
  {Urbakh}}]{PhysRevE.92.060101}%
  \BibitemOpen
  \bibfield  {author} {\bibinfo {author} {\bibfnamefont {T.}~\bibnamefont
  {Rotbart}}, \bibinfo {author} {\bibfnamefont {S.}~\bibnamefont {Reuveni}}, \
  and\ \bibinfo {author} {\bibfnamefont {M.}~\bibnamefont {Urbakh}},\
  }\href@noop {} {\bibfield  {journal} {\bibinfo  {journal} {Phys. Rev. E}\
  }\textbf {\bibinfo {volume} {92}},\ \bibinfo {pages} {060101} (\bibinfo
  {year} {2015})}\BibitemShut {NoStop}%
\bibitem [{\citenamefont {Majumdar}\ \emph {et~al.}(2015)\citenamefont
  {Majumdar}, \citenamefont {Sabhapandit},\ and\ \citenamefont
  {Schehr}}]{PhysRevE.92.052126}%
  \BibitemOpen
  \bibfield  {author} {\bibinfo {author} {\bibfnamefont {S.~N.}\ \bibnamefont
  {Majumdar}}, \bibinfo {author} {\bibfnamefont {S.}~\bibnamefont
  {Sabhapandit}}, \ and\ \bibinfo {author} {\bibfnamefont {G.}~\bibnamefont
  {Schehr}},\ }\href@noop {} {\bibfield  {journal} {\bibinfo  {journal} {Phys.
  Rev. E}\ }\textbf {\bibinfo {volume} {92}},\ \bibinfo {pages} {052126}
  (\bibinfo {year} {2015})}\BibitemShut {NoStop}%
\bibitem [{\citenamefont {Reuveni}(2016)}]{reuveni2016optimal}%
  \BibitemOpen
  \bibfield  {author} {\bibinfo {author} {\bibfnamefont {S.}~\bibnamefont
  {Reuveni}},\ }\href@noop {} {\bibfield  {journal} {\bibinfo  {journal} {Phys.
  Rev. Lett.}\ }\textbf {\bibinfo {volume} {116}},\ \bibinfo {pages} {170601}
  (\bibinfo {year} {2016})}\BibitemShut {NoStop}%
\bibitem [{\citenamefont {Pal}\ and\ \citenamefont
  {Reuveni}(2017)}]{pal2017first}%
  \BibitemOpen
  \bibfield  {author} {\bibinfo {author} {\bibfnamefont {A.}~\bibnamefont
  {Pal}}\ and\ \bibinfo {author} {\bibfnamefont {S.}~\bibnamefont {Reuveni}},\
  }\href@noop {} {\bibfield  {journal} {\bibinfo  {journal} {Phys. Rev. Lett.}\
  }\textbf {\bibinfo {volume} {118}},\ \bibinfo {pages} {030603} (\bibinfo
  {year} {2017})}\BibitemShut {NoStop}%
\bibitem [{\citenamefont {Belan}(2018)}]{belan2018restart}%
  \BibitemOpen
  \bibfield  {author} {\bibinfo {author} {\bibfnamefont {S.}~\bibnamefont
  {Belan}},\ }\href@noop {} {\bibfield  {journal} {\bibinfo  {journal} {Phys.
  Rev. Lett.}\ }\textbf {\bibinfo {volume} {120}},\ \bibinfo {pages} {080601}
  (\bibinfo {year} {2018})}\BibitemShut {NoStop}%
\bibitem [{\citenamefont {Bodrova}\ \emph {et~al.}(2019)\citenamefont
  {Bodrova}, \citenamefont {Chechkin},\ and\ \citenamefont
  {Sokolov}}]{bodrova2019nonrenewal}%
  \BibitemOpen
  \bibfield  {author} {\bibinfo {author} {\bibfnamefont {A.~S.}\ \bibnamefont
  {Bodrova}}, \bibinfo {author} {\bibfnamefont {A.~V.}\ \bibnamefont
  {Chechkin}}, \ and\ \bibinfo {author} {\bibfnamefont {I.~M.}\ \bibnamefont
  {Sokolov}},\ }\href@noop {} {\bibfield  {journal} {\bibinfo  {journal} {Phys.
  Rev. E}\ }\textbf {\bibinfo {volume} {100}},\ \bibinfo {pages} {012119}
  (\bibinfo {year} {2019})}\BibitemShut {NoStop}%
\bibitem [{\citenamefont {Ku\ifmmode~\acute{s}\else \'{s}\fi{}mierz}\ and\
  \citenamefont {Gudowska-Nowak}(2015)}]{PhysRevE.92.052127}%
  \BibitemOpen
  \bibfield  {author} {\bibinfo {author} {\bibfnamefont {L.}~\bibnamefont
  {Ku\ifmmode~\acute{s}\else \'{s}\fi{}mierz}}\ and\ \bibinfo {author}
  {\bibfnamefont {E.}~\bibnamefont {Gudowska-Nowak}},\ }\href@noop {}
  {\bibfield  {journal} {\bibinfo  {journal} {Phys. Rev. E}\ }\textbf {\bibinfo
  {volume} {92}},\ \bibinfo {pages} {052127} (\bibinfo {year}
  {2015})}\BibitemShut {NoStop}%
\bibitem [{\citenamefont {Campos}\ and\ \citenamefont
  {M\'endez}(2015)}]{PhysRevE.92.062115}%
  \BibitemOpen
  \bibfield  {author} {\bibinfo {author} {\bibfnamefont {D.}~\bibnamefont
  {Campos}}\ and\ \bibinfo {author} {\bibfnamefont {V.}~\bibnamefont
  {M\'endez}},\ }\href@noop {} {\bibfield  {journal} {\bibinfo  {journal}
  {Phys. Rev. E}\ }\textbf {\bibinfo {volume} {92}},\ \bibinfo {pages} {062115}
  (\bibinfo {year} {2015})}\BibitemShut {NoStop}%
\bibitem [{\citenamefont {Bhat}\ \emph {et~al.}(2016)\citenamefont {Bhat},
  \citenamefont {De~Bacco},\ and\ \citenamefont {Redner}}]{bhat2016stochastic}%
  \BibitemOpen
  \bibfield  {author} {\bibinfo {author} {\bibfnamefont {U.}~\bibnamefont
  {Bhat}}, \bibinfo {author} {\bibfnamefont {C.}~\bibnamefont {De~Bacco}}, \
  and\ \bibinfo {author} {\bibfnamefont {S.}~\bibnamefont {Redner}},\
  }\href@noop {} {\bibfield  {journal} {\bibinfo  {journal} {J. Stat. Mech.:
  Theory and Experiment}\ }\textbf {\bibinfo {volume} {2016}},\ \bibinfo
  {pages} {083401} (\bibinfo {year} {2016})}\BibitemShut {NoStop}%
\bibitem [{\citenamefont {Eule}\ and\ \citenamefont
  {Metzger}(2016)}]{eule2016non}%
  \BibitemOpen
  \bibfield  {author} {\bibinfo {author} {\bibfnamefont {S.}~\bibnamefont
  {Eule}}\ and\ \bibinfo {author} {\bibfnamefont {J.~J.}\ \bibnamefont
  {Metzger}},\ }\href@noop {} {\bibfield  {journal} {\bibinfo  {journal} {New
  J. Phys.}\ }\textbf {\bibinfo {volume} {18}},\ \bibinfo {pages} {033006}
  (\bibinfo {year} {2016})}\BibitemShut {NoStop}%
\bibitem [{\citenamefont {Ross}(2014)}]{ross2014introduction}%
  \BibitemOpen
  \bibfield  {author} {\bibinfo {author} {\bibfnamefont {S.~M.}\ \bibnamefont
  {Ross}},\ }\href@noop {} {\emph {\bibinfo {title} {Introduction to
  Probability Models}}}\ (\bibinfo  {publisher} {Academic Press},\ \bibinfo
{year} {2014})\BibitemShut {NoStop}%
\bibitem [{\citenamefont {Gnedenko}\ \emph {et~al.}(2014)\citenamefont
  {Gnedenko}, \citenamefont {Belyayev},\ and\ \citenamefont
  {Solovyev}}]{gnedenko2014mathematical}%
  \BibitemOpen
  \bibfield  {author} {\bibinfo {author} {\bibfnamefont {B.~V.}\ \bibnamefont
  {Gnedenko}}, \bibinfo {author} {\bibfnamefont {Y.~K.}\ \bibnamefont
  {Belyayev}}, \ and\ \bibinfo {author} {\bibfnamefont {A.~D.}\ \bibnamefont
  {Solovyev}},\ }\href@noop {} {\emph {\bibinfo {title} {Mathematical methods
  of reliability theory}}}\ (\bibinfo  {publisher} {Academic Press},\ \bibinfo
{year} {2014})\BibitemShut {NoStop}%
\bibitem [{\citenamefont {Feller}(1954)}]{feller1954diffusion}%
  \BibitemOpen
  \bibfield  {author} {\bibinfo {author} {\bibfnamefont {W.}~\bibnamefont
  {Feller}},\ }\href@noop {} {\bibfield  {journal} {\bibinfo  {journal} {Trans.
  Amer. Math. Soc.}\ }\textbf {\bibinfo {volume} {77}},\ \bibinfo {pages} {1}
(\bibinfo {year} {1954})}\BibitemShut {NoStop}%
\bibitem [{\citenamefont {Sherman}(1958)}]{sherman1958limiting}%
  \BibitemOpen
  \bibfield  {author} {\bibinfo {author} {\bibfnamefont {B.}~\bibnamefont
  {Sherman}},\ }\href@noop {} {\bibfield  {journal} {\bibinfo  {journal} {Ann.
  Math. Stat.}\ }\textbf {\bibinfo {volume} {29}},\ \bibinfo {pages} {267}
  (\bibinfo {year} {1958})}\BibitemShut {NoStop}%
\bibitem [{\citenamefont {Falcao}\ and\ \citenamefont
  {Evans}(2017)}]{falcao2017interacting}%
  \BibitemOpen
  \bibfield  {author} {\bibinfo {author} {\bibfnamefont {R.}~\bibnamefont
  {Falcao}}\ and\ \bibinfo {author} {\bibfnamefont {M.~R.}\ \bibnamefont
  {Evans}},\ }\href@noop {} {\bibfield  {journal} {\bibinfo  {journal} {J.
  Stat. Mech.: Theory and Experiment}\ }\textbf {\bibinfo {volume} {2017}},\
  \bibinfo {pages} {023204} (\bibinfo {year} {2017})}\BibitemShut {NoStop}%
\bibitem [{\citenamefont {{Ajjarapu}}\ \emph {et~al.}(1994)\citenamefont
  {{Ajjarapu}}, \citenamefont {{Ping Lin Lau}},\ and\ \citenamefont
  {{Battula}}}]{317656}%
  \BibitemOpen
  \bibfield  {author} {\bibinfo {author} {\bibfnamefont {V.}~\bibnamefont
  {{Ajjarapu}}}, \bibinfo {author} {\bibnamefont {{Ping Lin Lau}}}, \ and\
  \bibinfo {author} {\bibfnamefont {S.}~\bibnamefont {{Battula}}},\ }\href@noop
  {} {\bibfield  {journal} {\bibinfo  {journal} {IEEE Transactions on Power
  Systems}\ }\textbf {\bibinfo {volume} {9}},\ \bibinfo {pages} {906} (\bibinfo
  {year} {1994})}\BibitemShut {NoStop}%
\bibitem [{\citenamefont {{Zhihong Feng}}\ \emph {et~al.}(2000)\citenamefont
  {{Zhihong Feng}}, \citenamefont {{Ajjarapu}},\ and\ \citenamefont
  {{Maratukulam}}}]{867175}%
  \BibitemOpen
  \bibfield  {author} {\bibinfo {author} {\bibnamefont {{Zhihong Feng}}},
  \bibinfo {author} {\bibfnamefont {V.}~\bibnamefont {{Ajjarapu}}}, \ and\
  \bibinfo {author} {\bibfnamefont {D.~J.}\ \bibnamefont {{Maratukulam}}},\
  }\href@noop {} {\bibfield  {journal} {\bibinfo  {journal} {IEEE Transactions
  on Power Systems}\ }\textbf {\bibinfo {volume} {15}},\ \bibinfo {pages} {791}
  (\bibinfo {year} {2000})}\BibitemShut {NoStop}%
\bibitem [{\citenamefont {Miller}\ and\ \citenamefont
  {Orr}(1966)}]{miller1966model}%
  \BibitemOpen
  \bibfield  {author} {\bibinfo {author} {\bibfnamefont {M.~H.}\ \bibnamefont
  {Miller}}\ and\ \bibinfo {author} {\bibfnamefont {D.}~\bibnamefont {Orr}},\
  }\href@noop {} {\bibfield  {journal} {\bibinfo  {journal} {Quar. J. Econ.}\
  }\textbf {\bibinfo {volume} {80}},\ \bibinfo {pages} {413} (\bibinfo {year}
  {1966})}\BibitemShut {NoStop}%
\bibitem [{\citenamefont {Redner}(2001)}]{redner2001guide}%
  \BibitemOpen
  \bibfield  {author} {\bibinfo {author} {\bibfnamefont {S.}~\bibnamefont
  {Redner}},\ }\href@noop {} {\emph {\bibinfo {title} {A Guide to First-Passage
  Processes}}}\ (\bibinfo  {publisher} {Cambridge University Press},\ \bibinfo
  {year} {2001})\BibitemShut {NoStop}%
\bibitem [{\citenamefont {Feller}(2008)}]{feller2008introduction}%
  \BibitemOpen
  \bibfield  {author} {\bibinfo {author} {\bibfnamefont {W.}~\bibnamefont
  {Feller}},\ }\href@noop {} {\emph {\bibinfo {title} {An introduction to
  probability theory and its applications}}},\ Vol.~\bibinfo {volume} {1}\
  (\bibinfo  {publisher} {John Wiley \& Sons},\ \bibinfo {year}
  {2008})\BibitemShut {NoStop}%
\bibitem [{\citenamefont {Bray}\ \emph {et~al.}(2013)\citenamefont
  {Bray}, \citenamefont {Majumdar},\ and\ \citenamefont
  {Schehr}}]{Bray13}%
  \BibitemOpen
  \bibfield  {author} {\bibinfo {author} {\bibfnamefont {A.~J.}~\bibnamefont
  {Bray}}, \bibinfo {author} {\bibfnamefont {S.~N.}\ \bibnamefont
  {Majumdar}}, \ and\ \bibinfo {author} {\bibfnamefont {G.}~\bibnamefont
  {Schehr}},\ }\href@noop {} {\bibfield  {journal} {\bibinfo  {journal} {Adv. Phys.}\ }\textbf {\bibinfo {volume} {62}},\ \bibinfo {number} {3},\ \bibinfo {pages} {225}
  (\bibinfo {year} {2015})}\BibitemShut {NoStop}%
\bibitem [{\citenamefont {De~Bruyne}\ \emph {et~al.}(2021)\citenamefont
  {De~Bruyne}, \citenamefont {Redner},\ and\ \citenamefont
  {Randon-Furling}}]{BRR21}%
  \BibitemOpen
  \bibfield  {author} {\bibinfo {author} {\bibfnamefont {B.}~\bibnamefont
  {De~Bruyne}}, \bibinfo {author} {\bibfnamefont {S.}~\bibnamefont {Redner}}, \
  and\ \bibinfo {author} {\bibfnamefont {J.}~\bibnamefont {Randon-Furling}},\
  }\href@noop {} {\  (\bibinfo {year} {2021})}\BibitemShut {NoStop}%
\bibitem [{\citenamefont {Grigorescu}\ and\ \citenamefont
  {Kang}(2002)}]{grigorescu2002brownian}%
  \BibitemOpen
  \bibfield  {author} {\bibinfo {author} {\bibfnamefont {I.}~\bibnamefont
  {Grigorescu}}\ and\ \bibinfo {author} {\bibfnamefont {M.}~\bibnamefont
  {Kang}},\ }\href@noop {} {\bibfield  {journal} {\bibinfo  {journal} {Journal
  of Theoretical Probability}\ }\textbf {\bibinfo {volume} {15}},\ \bibinfo
  {pages} {817} (\bibinfo {year} {2002})}\BibitemShut {NoStop}%
\bibitem [{\citenamefont {Leung}\ and\ \citenamefont
  {Li}(2008)}]{leung2008spectral}%
  \BibitemOpen
  \bibfield  {author} {\bibinfo {author} {\bibfnamefont {Y.}~\bibnamefont
  {Leung}}\ and\ \bibinfo {author} {\bibfnamefont {W.}~\bibnamefont {Li}},\
  }\href@noop {} {\bibfield  {journal} {\bibinfo  {journal} {Proceedings of the
  American Mathematical Society}\ }\textbf {\bibinfo {volume} {136}},\ \bibinfo
  {pages} {4427} (\bibinfo {year} {2008})}\BibitemShut {NoStop}%
\bibitem [{\citenamefont {Ben-Ari}\ and\ \citenamefont
  {Pinsky}(2009)}]{ben2009ergodic}%
  \BibitemOpen
  \bibfield  {author} {\bibinfo {author} {\bibfnamefont {I.}~\bibnamefont
  {Ben-Ari}}\ and\ \bibinfo {author} {\bibfnamefont {R.~G.}\ \bibnamefont
  {Pinsky}},\ }\href@noop {} {\bibfield  {journal} {\bibinfo  {journal}
  {Stochastic processes and their applications}\ }\textbf {\bibinfo {volume}
  {119}},\ \bibinfo {pages} {864} (\bibinfo {year} {2009})}\BibitemShut
  {NoStop}%
\bibitem [{\citenamefont {Reuveni}\ \emph {et~al.}(2014)\citenamefont
  {Reuveni}, \citenamefont {Urbakh},\ and\ \citenamefont
  {Klafter}}]{RUR14}%
  \BibitemOpen
  \bibfield  {author} {\bibinfo {author} {\bibfnamefont {S.}~\bibnamefont
  {Reuveni}}, \bibinfo {author} {\bibfnamefont {M.}~\bibnamefont {Urbakh}}, \
  and\ \bibinfo {author} {\bibfnamefont {J.}~\bibnamefont {Klafter}},\
  }\href@noop {} {\bibfield  {journal} {\bibinfo  {journal} {PNAS}\ }\textbf {\bibinfo {volume} {111}},\ \bibinfo {number}
  {12},\ \bibinfo {pages}
  {4391} (\bibinfo {year} {2014})}\BibitemShut {NoStop}%
\bibitem [{\citenamefont {Evans}\ and\ \citenamefont
  {Majumdar}(2011{\natexlab{a}})}]{EMref18}%
  \BibitemOpen
  \bibfield  {author} {\bibinfo {author} {\bibfnamefont {M.~R.}\ \bibnamefont
  {Evans}}\ and\ \bibinfo {author} {\bibfnamefont {S.~N.}\ \bibnamefont
  {Majumdar}},\ }\href@noop {} {\bibfield  {journal} {\bibinfo  {journal}
  {J. Phys. A: Mathematical and Theoretical}\ }\textbf {\bibinfo {volume} {52}},\ \bibinfo {number} {1},\ \bibinfo {pages}
  {01LT01} (\bibinfo {year} {2018})}\BibitemShut {NoStop}%
\bibitem [{\citenamefont {Gallavotti}\ and\ \citenamefont
  {McKean}(1972)}]{gallavotti1972boundary}%
  \BibitemOpen
  \bibfield  {author} {\bibinfo {author} {\bibfnamefont {G.}~\bibnamefont
  {Gallavotti}}\ and\ \bibinfo {author} {\bibfnamefont {H.}~\bibnamefont
  {McKean}},\ }\href@noop {} {\bibfield  {journal} {\bibinfo  {journal} {Nagoya
  Mathematical Journal}\ }\textbf {\bibinfo {volume} {47}},\ \bibinfo {pages}
  {1} (\bibinfo {year} {1972})}\BibitemShut {NoStop}%
\bibitem [{\citenamefont {Harrison}\ and\ \citenamefont
  {Lemoine}(1981)}]{harrison1981sticky}%
  \BibitemOpen
  \bibfield  {author} {\bibinfo {author} {\bibfnamefont {J.~M.}\ \bibnamefont
  {Harrison}}\ and\ \bibinfo {author} {\bibfnamefont {A.~J.}\ \bibnamefont
  {Lemoine}},\ }\href@noop {} {\bibfield  {journal} {\bibinfo  {journal}
  {Journal of Applied Probability}\ }\textbf {\bibinfo {volume} {18}},\
  \bibinfo {pages} {216} (\bibinfo {year} {1981})}\BibitemShut {NoStop}%
\bibitem [{\citenamefont {Bou-Rabee}\ and\ \citenamefont
  {Holmes-Cerfon}(2020)}]{bou2020sticky}%
  \BibitemOpen
  \bibfield  {author} {\bibinfo {author} {\bibfnamefont {N.}~\bibnamefont
  {Bou-Rabee}}\ and\ \bibinfo {author} {\bibfnamefont {M.~C.}\ \bibnamefont
  {Holmes-Cerfon}},\ }\href@noop {} {\bibfield  {journal} {\bibinfo  {journal}
  {SIAM Review}\ }\textbf {\bibinfo {volume} {62}},\ \bibinfo {pages} {164}
  (\bibinfo {year} {2020})}\BibitemShut {NoStop}%
\bibitem [{\citenamefont {Rausand}\ and\ \citenamefont
  {H{\o}yland}(2003)}]{rausand2003system}%
  \BibitemOpen
  \bibfield  {author} {\bibinfo {author} {\bibfnamefont {M.}~\bibnamefont
  {Rausand}}\ and\ \bibinfo {author} {\bibfnamefont {A.}~\bibnamefont
  {H{\o}yland}},\ }\href@noop {} {\emph {\bibinfo {title} {System reliability
  theory: models, statistical methods, and applications}}},\ Vol.\ \bibinfo
  {volume} {396}\ (\bibinfo  {publisher} {John Wiley \& Sons},\ \bibinfo {year}
  {2003})\BibitemShut {NoStop}%
\bibitem [{\citenamefont {Harrison}\ \emph {et~al.}(1983)\citenamefont
  {Harrison}, \citenamefont {Sellke},\ and\ \citenamefont
  {Taylor}}]{harrison1983impulse}%
  \BibitemOpen
  \bibfield  {author} {\bibinfo {author} {\bibfnamefont {J.~M.}\ \bibnamefont
  {Harrison}}, \bibinfo {author} {\bibfnamefont {T.~M.}\ \bibnamefont
  {Sellke}}, \ and\ \bibinfo {author} {\bibfnamefont {A.~J.}\ \bibnamefont
  {Taylor}},\ }\href@noop {} {\bibfield  {journal} {\bibinfo  {journal}
  {Mathematics of Operations Research}\ }\textbf {\bibinfo {volume} {8}},\
  \bibinfo {pages} {454} (\bibinfo {year} {1983})}\BibitemShut {NoStop}%
\bibitem [{\citenamefont {Buckholtz}\ \emph {et~al.}(1983)\citenamefont
  {Buckholtz}, \citenamefont {Campbell}, \citenamefont {Milbourne},\ and\
  \citenamefont {Wasan}}]{buckholtz1983analysis}%
  \BibitemOpen
  \bibfield  {author} {\bibinfo {author} {\bibfnamefont {P.~G.}\ \bibnamefont
  {Buckholtz}}, \bibinfo {author} {\bibfnamefont {L.~L.}\ \bibnamefont
  {Campbell}}, \bibinfo {author} {\bibfnamefont {R.~D.}\ \bibnamefont
  {Milbourne}}, \ and\ \bibinfo {author} {\bibfnamefont {M.}~\bibnamefont
  {Wasan}},\ }\href@noop {} {\bibfield  {journal} {\bibinfo  {journal} {Journal
  of Applied Probability}\ }\textbf {\bibinfo {volume} {20}},\ \bibinfo {pages}
  {61} (\bibinfo {year} {1983})}\BibitemShut {NoStop}%
\bibitem [{\citenamefont {Tavar{\'e}}(1979)}]{tavare1979dual}%
  \BibitemOpen
  \bibfield  {author} {\bibinfo {author} {\bibfnamefont {S.}~\bibnamefont
  {Tavar{\'e}}},\ }\href@noop {} {\bibfield  {journal} {\bibinfo  {journal}
  {Theoretical population biology}\ }\textbf {\bibinfo {volume} {16}},\
  \bibinfo {pages} {253} (\bibinfo {year} {1979})}\BibitemShut {NoStop}%
\bibitem [{\citenamefont {Goel}\ and\ \citenamefont
  {Richter-Dyn}(2016)}]{goel2016stochastic}%
  \BibitemOpen
  \bibfield  {author} {\bibinfo {author} {\bibfnamefont {N.~S.}\ \bibnamefont
  {Goel}}\ and\ \bibinfo {author} {\bibfnamefont {N.}~\bibnamefont
  {Richter-Dyn}},\ }\href@noop {} {\emph {\bibinfo {title} {Stochastic models
  in biology}}}\ (\bibinfo  {publisher} {Elsevier},\ \bibinfo {year}
  {2016})\BibitemShut {NoStop}%
\end{thebibliography}

%

\end{document}